\documentclass[twocolumn,pra,eqsecnum,keywords]{revtex4}

\usepackage{bm}
\usepackage{amsmath}
\usepackage{graphicx}

\begin{document}
\date{\today}
\title{Fully QED/relativistic theory of light pressure on free electrons by isotropic radiation}
\vspace{-.5in}
\author{A. E. Kaplan}
\affiliation{Electr. \& Computer Engineering Dept., Johns Hopkins University, Baltimore, 
MD, 21218, USA}\email{alexander.kaplan@jhu.edu}
\begin{abstract}
A relativistic/QED theory of light pressure on electrons 
by an isotropic, in particular blackbody
radiation predicts thermalization rates of 
free electrons over entire span of 
energies available in the lab and the nature.
The calculations based on 
the QED Klein-Nishina theory of electron-photon scattering
and relativistic Fokker-Planck equation,
show that the transition from
classical (Thompson) to QED (Compton) thermalization
determined by the product of electron energy
and radiation temperature, is reachable 
under conditions for controlled nuclear fusion,
and predict large acceleration of 
electron thermalization in the Compton domain
and strong damping of plasma oscillations
at the temperatures near plasma nuclear fusion.
\end{abstract}
\maketitle 
\section{Introduction}
Beginning with Max Planck discoveries [1],
one of the fundamental issues 
in optics, electrodynamics, thermodynamics, atomic physics 
and quantum mechanics, is how a radiation, in particular blackbody
radiation, imposes an equilibrium in the material system 
by either heating it up or cooling down.
This process is facilitated by
a so called light pressure [2-4] on charged particles [5],
most of all the lightest ones -- electrons,
either (quasi)free as in plasma, of bound as in atoms or ions,
in which case the electrons pass the light pressure on to atoms.
The advent of lasers allowed the development of
highly controlled and engineered non-thermal
radiation environment, such as coherent laser light with
its frequency tuned near atomic resonances,
and use them for the cooling of atoms by red-shifted laser [6-9]
(and coherent motional excitation of atoms by blue-shifted one [10,11]).
A related ponderomotive, or field-gradient force [12-14],
manifested e. g. $via$ Kapitza-Dirac effect [15-18],
has also been used in 
laser trapping of atoms [6-9] and macro-particles [19],
high-field ionization of atoms [19-21], etc.
It can be very sensitive to relativistic effects,
which under certain conditions
may result in a chaotic motion [22,23] or even
reverse the sign of that force in a strong field [24,25].

In the case of blackbody
radiation acting upon free electrons, 
the situation is rather straightforward and 
fundamental: non-resonant light pressure 
plays the role of the equilibrium 
"enforcer" by either energizing slow electrons 
or damping the fast ones.
The main issue here is how fast 
the equilibrium/thermalization can be reached.
The relaxation time, $t_{rlx}$, of that process
could vary by many orders
of magnitude depending of the temperature
$T$ and initial energy/momentum of electrons;
in classical domain $t_{rlx} \propto T^{-4}$.
At the temperatures near absolute zero 
(e. g. in the so called relic radiations,
or Cosmic Microwave Background, CMB [26-28])
the equilibrium is essentially unreachable 
(it is too long even at $T \sim 10^3 K$, see below),
whereas in high-$T$ environments,
e. g. controlled nuclear fusion,
nuclear explosions, and star cores, 
it can be reached faster than in attoseconds.
We found that the nature of transition to equilibrium
is controlled by a parameter
called by us "Compton factor",
$K_C = q \theta \gamma $,
where $\gamma$ is a relativistic 
factor of electrons,
$\theta = k_B T / m_0 c^2$ is a normalized temperature, 
$k_B$ is the Boltzmann constant, 
$m_0 c^2$ -- the rest energy of electron, and
$q \approx 10$, see below.
When $K_C \ll 1$,
we are dealing with a so called Thompson,
or classical, scattering of light,
whereby the scattering cross-section,
$\sigma_0$ is constant, even if $\gamma \gg 1$,
and the theory of light pressure 
is well known, see e. g. [5].
Due to advents in laser and 
controlled fusion technologies,
the $K_C$ can be large enough, $K_C > 1$,
and we are entering a QED, or Compton domain,
where the respective theory is far less developed.

In view of new developments in nuclear fusion 
and physics in general,
e. g. in astrophysics and cosmology [29-31],
it would be of fundamental importance
to have the theory of that force
over all the energies of electrons
and temperatures of radiation,
from the classical, $K_C \ll 1$, to transient,
$K_C \sim 1$, to QED domains, $K_C \gg 1$.
From QM viewpoint [32,33], the light pressure on elementary
particles is a result of averaging over ensemble of
Inverse Compton Scattering events [34-36],
whereby a particle transfers part
of its momentum to a scattered photon.
The known results provide 
patchy descriptions of the process,
with quantitative results known mostly for 
``cold'' case [5],
and qualitative -- for very ``hot'' case 
in the theory of high-energy cosmic rays [36-38].
Yet, to the best of our knowledge,
no general formula for the 
light pressure $F ( \gamma, \theta )$
and related relaxation rates
for arbitrary $K_C$ presently exists.

In this paper, 

(a) by using Lorentz transformation of spectrum, Sect. II,
and relating photon scattering to momentum transfer
to electrons $via$ Doppler and Compton relationships, Sect. III,
we derived a general light pressure formula, Sect. IV,
for the isotropic/uniform radiation
with an $arbitary$ frequency spectrum,
$arbitaryly$ relativistic energy of electrons
and $arbirary$ spectral dependence of the
the scattering cross-section on frequency of light, 

(b) checked it out for the well known
case of Thompson scattering
and related light pressure, Sect. V,
and simplified it in the case of
blackbody radiation with Planks spectrum at
$arbitary$ temperature, Sect. VI;

(c) by using the frequency/energy dependence
of photon-electron scattering cross-section in the Compton, or QED
domain, based on the Klein-Nishina QED theory [39,40] which
takes into consideration virtual electron-positron pair 
creation and annihilation, Sect. VII,
we applied a general formula to the case of electron,
immersed in the blackbody radiation,  
and found amazingly precise and universal analytic approximation
for the light force in the entire domain, Sect. VIII, and finally

(d) considered kinetics of electron density distribution
to the equilibrium, i. e. the thermalization of the distribution,
using Fokker-Planck equation and its solutions,
in particular relaxation rates of the process,
and how they may affect plasma oscillations, Sect. IX.

Our results may have important
applications for both high-temperature plasma,
in particular controlled nuclear fusion
and nuclear explosions,
and to astrophysics/cosmology
(to be addressed  by us elsewhere [41]).
\vspace {-.2in}
\section{Lorentz transformation of radiation spectrum}
An isotropic (and homogeneous) radiation 
is associated with some preferred $L$-frame, where
a particle at rest experiences 
no time averaged light pressure,
as the action of a $\vec{k} $-vector component
is canceled by a counter-propagating ($ - \vec{k} $) component.
We assume the spectrum of this radiation,
$\rho_{_L} ( \omega )$, known.
A particle moves in the $x$-axis in that frame
with velocity $\vec{v} = $ $v \hat{e}_x$,
and is at rest in a certain $P$-frame.
A light pressure $ \vec{F} = d \vec{p} / dt = F \hat{e}_x$
on a particle is then nonzero if $v \neq 0$;
here $\vec{p} = p \hat{e}_x$ is its momentum in the $L$-frame,
$p / m_0 c \equiv \mu = \beta \gamma = \sqrt { \gamma^2 - 1}$,
where $\beta = $ $v / c$, $m_{_0}$ is a particle rest mass,
$c$ -- speed of light, and
$\gamma =$ $ 1 / \sqrt { 1 - \beta^2 } =$ $ \sqrt { 1 + \mu^2 }$.
Our derivation of $F ( \mu )$ is
based on Lorentz transformation of 
$\rho ( \vec k )$ from $L$-frame to $P$-frame [42,43];
it is valid for arbitrary frequency
dependence of a full cross-section $\sigma ( \epsilon )$
where $\epsilon = \hbar  \omega / m_0 c^2$,
of scattering of an $\omega$-photon at a particle.

For a gas of particles with non-zero mass,
one can define distribution function $g ( \vec{p} , \vec{r} )$ in
a lab $L$-frame in the phase space of momentum $\vec{p}$
and position vector $\vec{r}$ as the number of particles, 
$dN_{\vec{p} , \vec{r} } = g ( \vec{p} , \vec{r} )  d \Omega$,
per the element of phase space,
$d \Omega = dV_{\vec{p}} dV_{\vec{r}}$, 
where $dV_{\vec{p}} = dp_x dp_y dp_z$
and $dV_{\vec{r}} = dx \ dy \ dz$ are the elements of momentum and
coordinate spaces respectively.
A general formula for a Lorentz transformation of a
distribution function $g_{_L} ( \vec{p}_{_L} , \vec{r}_{_L} )$ in
the $L$-frame to a distribution function 
$g_{_P} ( \vec{p}_{_P} , \vec{r}_{_P})$
in a $P$-frame moving uniformly with respect to
the $L$-frame is as [42,43]:
\begin{equation}
g_{_L} ( \vec{p}_{_L} , \vec{r}_{_L} ) = 
g_{_P} ( \vec{p}_{_P} , \vec{r}_{_P} ) ;
\ \ \ \ \ 
d \Omega_{_L} = d \Omega_{_P}
\tag{1}
\end{equation}
where $\vec{k }_{_L}$ and $\vec{r}_{_L}$
are related to $\vec{k}_{_P}$ and $\vec{r}_{_P}$
respectively by a standard Lorentz transform for
an observer moving in $L$-frame in the
$x$-axis with velocity $\vec{v} = \hat{e}_x v$.
In the case of photon gas Eq. (1) remains true 
for the spectrum of photons $g ( \vec{k} , \vec{r})$,
by replacing $\vec{p }$ with
$\vec{k}$ ($k = \omega / c$),
$dV_{\vec{p}}$ by $dV_{\vec{k}} = dk_x dk_y dk_z$,
and $d \Omega$ by $d \tilde{\Omega} = dV_{\vec{k}} dV_{\vec{r}}$.
Since we are interested here in the case 
of a homogeneous radiation, a spectrum is $g ( \vec{k} ) $, 
and its transformation is written as
\begin{equation}
g_{_L} ( \vec{k }_{_L} ) = g_{_P} ( \vec{k}_{_P}) ,
\ \ \ \ \ 
d \tilde{\Omega}_{_L} = d \tilde{\Omega}_{_P} ; \ \ \ \
\tag{2}
\end{equation}
(here $g$ and $d \tilde{\Omega}$ - dimensionless),
where Lorentz transform for $\vec{k }$ 
is $( k_x )_{_L} = \gamma [ ( k_x )_{_P} + \beta k_{_P} ]$,
$( \vec{k}_\bot )_{_L} = ( \vec{k}_\bot )_{_P}$;
$k_{_L} = \gamma [ k_{_P} + \beta ( k_x )_{_P} ]$,
with $\vec{k}_\bot = k_y \hat{e_y}$ $ + k_z \hat{e_z} $,
$k = \omega / c = \sqrt { k_x^2 + k_\bot^2 }$, 
In particular, the Doppler coefficient is as
\begin{equation}
D \equiv {\omega_{_P}} / {\omega_{_L}} = 
[ \gamma ( 1 + \beta \cos \xi_{_P} ) ]^{- 1} =
\gamma ( 1 - \beta \cos \xi_{_L} ) 
\tag{3}
\end{equation}
where $ \xi_{(...)}$ are the angles
between respective $\vec{k}$-vectors 
and the $x$-axis, i.e.  $\cos \xi_{_L} = ( k_x )_{_L} / k_{_L}$,
$\cos \xi_{_P} = ( k_x )_{_P} / k_{_P}$,
transformed as
$\cos \xi_{_P} = ( \cos \xi_{_L} - \beta ) / ( 1 - \beta \cos \xi_{_L} )$.
Furthermore, since the radiation 
is isotropic in the $L$-frame,
the distribution function $g_{_L}$ does not depend
on the direction of $\vec{k_{_L}}$-vector, and
we have $g_{_L} ( \vec{k }_{_L} ) = g_{_L} ( {k }_{_L} )$,
where $k = | \vec{k}|$.
Since both spectra are symmetrical around the $x$-axis, 
we will use spherical coordinates in the $\vec{k}$ space,
so that $dV_{\vec{k}} = k^2 dk \ dO$, where $dO$ 
is the element of solid angle in the direction of $\vec{k}$. 
One can then introduce 
the density number of photons in the element 
$\rho ( \vec{k} )$,
$dk \ dO V_{\vec{r}}$ defined as
\begin{equation}
\rho ( \vec{k} ) = k^2 g ( \vec{k} )  \ \ \
with \ \ \
dN_{\vec{k} } = \rho ( \vec{k} )  dk \ dO \ dV_{\vec{r}} ;
 \ \ \ \ \ 
\tag{4}
\end{equation}
($[ \rho ] = cm^{-2}$),
and its transformation using Eq. (2) as:
\begin{equation}
\rho_{_L} ( {k}_{_L} ) / k_{_L}^2 = \rho_{_P} ( \vec{k}_P ) / k_{_P}^2 
\tag{5}
\end{equation}
From now on, since we are dealing 
with a homogeneous radiation independent on 
the radius-vector length, $| \vec{r} |$,
we will re-assign the notion of spectra $\rho$ only 
to the ones being functions 
of frequency $\omega$ and angles $\xi$
instead of $\vec{k}$ and $\vec{r}$ vectors.
In this case, $\rho ( \omega , \xi ) d \omega$
will have dimension of $[cm^{-3}]$.
As expected, the radiation becomes anisotropic in the $P$-frame
(yet symmetric around the $x$-axis)
with its spectrum $\rho_{_P} ( \omega , \xi_{_P} )$
given by a $L$-frame isotropic spectrum 
$\rho_{_L} ( \omega )$, whose argument and
amplitude are now altered 
by the parameters $\beta$ and $\xi_{_P}$
$via$ Doppler coefficient 
$D = [ \gamma ( 1 + \beta \cos \xi_{_P} ) ]^{-1}$, Eq. (3): 
\begin{equation}
\rho_{_P} ( \omega , \xi_{_P} ) = \rho_{_L} 
[ \omega / D (  \beta , \xi_{_P} ) ]  D^2 (  \beta , \xi_{_P} )
\tag{6}
\end{equation}
Our further calculations will mostly be focussed on the events
in the $P$-frame assuming that 
$\rho_{_L} ( \omega )$ is known.
\vspace{-0.2in}
\section{Photon scattering and momentum transfer}
We designate the wave-vector of an incident photon in
the $P$-frame as $\vec{k}_{in}$, 
and that of a scattered photon as $\vec{k}_{sc}$.
The latter one is scattered into the solid angle
$dO_{sc} = \sin \psi_{sc} d \psi_{sc} d \phi$
around $\vec{k}_{sc}$.
Here $\psi_{sc} \in [ 0 ,  \pi )$
is the angle between incident and scattered $k$-vectors,
and $\phi \in [ 0 , 2 \pi )$
is an azimuthal angle around the $\vec{k}_{sc}$ direction.
The number of photons scattered
into $dO_{sc}$ within time interval $dt$
and spectral band $d \omega = c  dk$, is
\begin{equation}
dN_{sc} = \rho_{_P} ( \omega_{in} , \xi_{in} ) ( d \sigma /  dO_{sc} )
( dO_{sc} / 4 \pi ) c dt d \omega
\tag{7}
\end{equation}
where the differential 
scattering cross-section $d \sigma /  dO_{sc}$
describes an (unknown yet)
physics of energy and momentum transfer
from the incident photon to
both scattered photon and a particle
[including a possible excitation
of internal degrees of freedom
in the particle if there is any,
which is not the case for an electron,
whereby $d \sigma /  dO_{sc}$ is
strictly due to Compton
scattering, see next equation (8),
that enters Klein-Nishina formula, 
see below, Eq. (27), and discussion 
in the end of Sect. IV].

The Compton quantum formula determines
the ratio $R$ of photon energies after and before scattering
from a single electron:
\begin{equation}
\epsilon_{sc}/\epsilon_{in} =
R ( \epsilon_{in} , \psi_{sc} ) =
[ {1 + \epsilon_{in} ( 1 - \cos \psi_{sc} ) } ]^{-1}
\tag{8}
\end{equation}
where $\epsilon \equiv \hbar \omega / m_0 c^2 $.
When calculated back to the $L$-frame,
photons back-scattered from an ultra-relativistic
electron, $\gamma \gg 1$,
may have large energies with the Doppler shift
up to $D \approx 2 \gamma$,
even if their energies in the $P$-frame
are still below QED limit, i. e.  $\epsilon_{in} \ll 1$.
It is commonly called an Inverse Compton 
(or Thompson, if $\epsilon_{in} \ll 1$) scattering.

Only the projection of $\vec{k}_{sc}$
into $\vec{k}_{in}$-direction
contribute to the force $F$;
all the rest are canceled out after 
integration over the azimuthal angle $\phi$
(in the $P$-frame, where the electron is at rest, 
the scattering problem has a symmetry around $\vec{k}_{in}$).
Thus the momentum transfer to an electron in 
$\vec{k}_{in}$-direction after the scattering is 
\begin{equation}
\hbar \Delta k_{tr} =
\hbar ( {k}_{in} - {k}_{sc} \cos \psi_{sc}) =
\hbar  {\omega}_{in} [ 1 - R ( \epsilon , \psi_{sc} ) \cos \psi_{sc} ] / c
\tag{9}
\end{equation}
\vspace{-0.3in}
\section{Light pressure on a particle}
Considering the number of photons, $d N_{sc}$, Eq. (7),
scattered into a solid angle $dO_{sc}$,
the light pressure impacted by them on the electron in
$\vec{k}_{in}$-direction, 
is the rate of momentum transfer,
$\vec{\mathcal {F}} = \mathcal {F} \vec{q}_{in}$
where
$ \vec{q}_{in} = \vec{k}_{in} / {k}_{in}$ and
\begin{equation}
d \mathcal {F} = \hbar \Delta k_{tr}
\frac { d^2 N_{sc} } {dt d \omega_{in} } =
\frac {\hbar {\omega}_{in}} {4 \pi}
\frac {d \sigma} {dO_{sc}}  \times
\notag
\end{equation}
\vspace{-.2in}
\begin{equation}
[ 1 - R ( \epsilon, \psi_{sc} ) \cos \psi_{sc} ] 
\rho_{_P} ( \omega_{in} , \xi_{in} ) \sin \psi_{sc}
d \psi_{sc} d \phi 
\tag{10}
\end{equation}
Integrating Eq. (10) over $\phi$ and $\psi_{sc}$,
we find a full $\omega_{in}$-Fourier component of the light pressure as:
\vspace{-.05in}
\begin{equation}
\mathcal {F} ( \omega_{in} , \beta ,  \vec{q}_{in} ) = 
( \hbar {\omega}_{in} / 4 \pi )
\rho_{_P} ( \omega_{in} , \xi_{in} ) \sigma_{_{MT}}  
\tag{11}
\end{equation}
\vspace{-.1in}
with
\begin{equation}
\sigma_{_{MT}} =
2 \pi \int_0^\pi [ 1 - R ( \epsilon, \psi_{sc} ) \cos \psi_{sc} ] 
\frac{d \sigma}{dO_{sc}} \sin \psi_{sc} d \psi_{sc}
\tag{12}
\end{equation}
where $\sigma_{_{MT}}$ is the {\it cross-section 
of a momentum transfer}
from $\omega_{in}$-photons to a particle.
It must be noted that $\sigma_{_{MT}}$
$\emph {may not}$ in general coincide with
a plain $full$ (integrated)
scattering cross-section,  $\sigma ( \epsilon_{in} )$
\begin{equation}
\sigma ( \epsilon_{in} )= 
2 \pi \int_0^{\pi} \frac{d \sigma ( \epsilon_{in} ) }
{dO_{sc}} \sin \psi_{sc} d \psi_{sc}; \ \ \
\epsilon_{in} \equiv \frac {\hbar \omega_{in}} {m_0 c^2}
\tag{13}
\end{equation}
because only the projection of $\vec{k}_{sc}$
into $\vec{k}_{in}$-direction
contributes to the radiation force $F$, see Eqs. (9) and (11),
where in general 
$R ( \epsilon, \psi_{sc} ) < 1$ if $\epsilon > 0$, Eq. (8).
$\sigma_{_{MT}}$ and $\sigma ( \epsilon_{in} )$ are related as
$\sigma_{_{MT}} = \sigma - \sigma_R$, where
\begin{equation}
\sigma_{_R} ( \epsilon ) = 
2 \pi \int_0^{\pi} R ( \epsilon, \psi ) 
\frac{d \sigma}{dO} \cos \psi \sin \psi d \psi =
\notag
\end{equation}
\vspace{-.1in}
\begin{equation}
2 \pi \int_{-1}^1 \frac{d \sigma}{dO}
\frac{\zeta d \zeta}{1 + \epsilon ( 1 - \zeta )}
\tag{14}
\end{equation}
which in turn reflects $R$-factor in Eq. (9) and (12).
It zeroes out for $\sigma = const$
(Thompson scattering) and at $\epsilon = 0$,
and peaks at $\epsilon \sim 0.54$ (see the end of Sect. VII),
but is negligibly small both at $\epsilon \ll 1$ and $\epsilon \gg 1$.

Now, we compute the light pressure,
$\vec{F} = F ( p ) \hat{e}_x$,
as an integral of the $x$-projections of the Fourier
force components, 
$\mathcal {F} ( \omega_{in} , \beta , \vec{q}_{in} )$, i. e.
$\mathcal {F}_x = \vec{\mathcal {F}} ( \omega_{in} , \beta , 
\vec{q}_{in} ) \hat{e}_x =
\mathcal {F} \cos \xi_{in}$, over all the incident solid angles,
$dO_{in} = \sin \xi_{in} d \xi_{in} d \phi_{in}$, 
and frequencies $\omega_{in}$ in the $P$-frame.
Thus, we have for the full light pressure
\begin{equation}
F ( p ) \equiv \frac{dp}{dt} =
\int \int \frac{d\mathcal {F}_x}{dO_{in}} dO_{in} d\omega =
\notag
\end{equation}
\vspace{-.2in}
\begin{equation}
2 \pi  \int_0^{\infty} \left[ \int_0^{\pi}  
\mathcal {F} ( \omega_{in} , \beta , \xi_{in} )
\sin \xi_{in} \cos \xi_{in} d \xi_{in} \right] d\omega
\tag{15}
\end{equation}
Recalling that a spectrum $\rho_{_P}$ in 
$\mathcal {F}$,  Eqs. (11),(15) 
can be expressed directly $via$ the known isotropic spectrum 
in the $L$-frame, $\rho_{_L}$, Eq. (6),
we can now write the expression for
the light pressure in closed form as:
\begin{equation}
F ( p ) = \frac{\hbar}{4}
\int_0^{\infty} 
\omega \sigma_{_{MT}} ( \omega )
\left[ \int_0^{\pi} {\sin (2 \xi ) \rho_{_L} ( \omega / D ) }
{D^2} d \xi \right] d \omega  \ \ \ \ \
\tag{16}
\end{equation}
or by  using a substitution, $\zeta = \cos \xi$, reduce it to:
\begin{equation}
F ( p ) = \frac {\hbar} 2 \int_0^{\infty} 
\frac{\omega \sigma_{_{MT}} ( \omega )}{ \gamma^2} 
\left\{ \int_{-1}^{1} \frac{\rho_{_L} [ \omega \gamma 
( 1 + \beta \zeta ) ] }
{( 1 + \beta \zeta )^2}
\zeta d \zeta \right\} d \omega 
\tag{17}
\end{equation}
Alternatively, by using a substitute
$\nu = \omega \gamma ( 1 + \beta \zeta )$,
the same result can be written as
\begin{equation}
F(p) = \frac {\hbar} 2 \int_0^{\infty} 
\frac{\nu \rho_{_L} ( \nu )}{\gamma^4 } 
\left[ \int_{-1}^{1} \sigma_{_{MT}} 
\left( \frac{\nu / \gamma}{1 + \beta \zeta } \right)
\frac{ \zeta d \zeta }{ ( 1 + \beta \zeta )^4 }
\right] d \nu  \ \
\tag{18}
\end{equation}
Note also that only $full$ (integrated) cross-section, 
$\sigma_{_{MT}}$, enters into the final calculations.
Which one of Eqs. (17) or (18)
to use for detailed
study is a matter of computational
convenience depending on specific
model functions $\sigma_{_{MT}} ( s )$
and $\rho_{_L} ( \nu )$
(see Sect. VI below for Planck radiation);
both of them  incorporate ensemble averaging over
all the relevant parameters, which
makes pressure $F$ the best tool to 
explore electron de-acceleration in EM field.

Let us reflect on the domain 
of validity of Eqs. (17) and (18).
The rest of this paper is dealing
with the light pressure on a single,
most fundamental elementary particle, electron
(or substantially rarefied electron gas or plasma),
which presents a clear case.
The question is then whether they could be
applicable for more general cases,
in particular for high density gas or plasma with
many-body interactions, in particular plasma oscillations, 
and for single particles/objects with
an internal structure and resonances.
We address the former issue in the Sect. IX
below [see the text preceding and following Eq. (39)],
and discuss the latter one here.

In our derivation of Eqs. (17), (18) 
we have not used any assumption based on the fact
that a scattering object is an elementary particle.
Essentially, the only assumption was 
that we have only one particle and one photon
both in input and final output channels
in each act of scattering, and were not concerned
about intermediate processes.
The physics of these processes
in general is to be described
by the differential cross-section 
$d \sigma ( \omega ) / d O_{sc}$, which in the case
of an electron is due to
Klein-Nishina formula, see below Eq. (27),
based on Compton scattering relationship
for input/output electron energies, Eq. (8);
eventually, $d \sigma / d O_{sc}$ is
absorbed into the full cross-sections
of scattering $\sigma ( \omega )$
and momentum transfer, $\sigma_{_{MT}} ( \omega )$
$via$ integration, Eq. (13) and Eq. (12),
which may also include a new ratio R of photon 
energies after and before scattering from a particle, 
that may differ now from the Compton quantum formula, 
Eq. (8), for a single electron, 
based now on radiation loss or amplifications 
after scattering from the particle 
due to its internal degrees of freedom.
Thus, Eqs. (17), (18)
use only our knowledge of $\sigma$ and
$\rho$ as functions of $\omega$.
They will be valid for example for
the resonant or other dispersion-related
interaction of the radiation with atoms or
even macro-particles, with any quantum or classical
resonances due to e. g. dipole momenta,
band structure, eigen-modes, etc.
By the same token it also does not matter whether
the cross-section $\sigma$ is due to
elastic scattering or includes losses
of energy to internal degrees of freedom;
all that implicitly enters into the functions
$\sigma ( \omega )$ and $\sigma_{_{MT}} ( \omega )$.
To a degree, this is reminiscent of a phenomenological
role played by dispersive dielectric 
constant $\varepsilon ( \omega )$
in electrodynamics whereby
$\varepsilon ( \omega )$ provides a short-hand
representation of all the constitutive interactions
of EM-field with matter.

The above discussion clearly suggests
the situations whereby the description
offered by Eqs. (17), (18) is not comprehensive:
it is when there are more than one output particle
(as e. g.  in the case of photoionization resulting
in an ion and one or more ionized electrons),
and/or more than one output photon
(as e. g. in stimulated emission
or any kind of nonlinear multi-photon process,
such as sum, difference, or 
high harmonics generation, etc).
In all those cases, one needs to 
include all the scattering channels
and generalize Eqs. (17), (18) 
by summation over all of them.
\vspace{-.1in}
\section{Thompson light pressure ($\bf {\sigma_{_{MT}} =  \sigma = const}$, $\sigma_{_R} = 0$ )}
\vspace{-.1in}
In the limit of frequency-independent
cross-section, the integral in Eq. (17)
is readily evaluated resulting in:
\begin{equation}
F_{_{Th}} = - (4/3) \sigma_0 W_{_L} \mu \gamma , \ \ with \ \
\gamma = \sqrt { 1 + \mu^2 }
\tag{19}
\end{equation}
where $W_{_{L}} = \hbar \int_0^\infty \omega \rho_{_L} ( \omega ) d \omega$
is the energy density of photons in the $L$-frame,
and $\sigma_0$ is a full (classical) cross-section 
of a charged particle:
\begin{equation}
\sigma = const =
\sigma_{_0} = ({ 8 \pi } / 3 )  r_{_0}^2 ;
\ \ \ with \ \ \
r_{_0}  = e^2 / { m_{_0} c^2 }
\tag{20}
\end{equation}
where $r_{_0}$ is  a classical EM-radius of a particle.
Eq. (19) coincides with known results [5].
At $\mu \ll 1$ we have $- F \propto \mu$, 
while in a relativistic case, $\mu \gg 1$, 
$- F \propto$ $ \mu | \mu |$.
It is reminiscent of a drag force in liquids and gases,
which is linear in velocity $v$ for low $v$ (Stokes force),
and $\propto v | v |$ for highly turbulent flow [44].
\section{Light pressure of a blackbody (Planck) radiation}
Spectral, $\rho_{_{TR}}$, 
and total energy, $W_{_{TR}} =
\hbar \int_0^\infty \omega \rho_{_{TR}} ( \omega )$ $ d \omega$
densities of blackbody radiation in the $L$-frame  at the temperature 
$T$ ($\approx 2.725 K$ for current CMB) are
\vspace {-.09in}
\begin{equation}
\rho_{_{TR}} ( \omega ) d \omega  = \frac {\omega^2}
{\pi^2 c^3 } \frac {d \omega}  {e^{ \hbar \omega / k_{_B} T }  - 1};
\ \ \
W_{_{TR}} = \frac{8 \pi^5}{15} W_{_C} \theta^4 
\tag{21} 
\end{equation} 
which is a familiar Planck density distribution, where 
$k_{_B}$ is the Boltzmann constant,
$W_{_C} = m_{_0} c^2  /  \lambda_{_C}^3 $
is a ``Compton energy density'',
$\lambda_{_C} = 2 \pi \hbar / m_0 c$
is the Compton wavelength,
and $\theta = k_{_B} T /  m_{_0} c^2$
is a dimensionless temperature
(for the current CMB,
$\theta \approx 0.534 \times$ $10^{-9}$).
In Thompson limit, 
the energy density $W_{_L}$ can now be replaced 
by $W_{_{TR}}$. 
For further calculations, we will
use a dimensionless time $\tau = t / t_{_C}$, and
force $f = F t_{_C} / m_0 c$ by introducing
a ``Compton time scale'' for an electron:
\begin{equation}
t_{_C} = 135 \lambda_{_C} / 64 \pi ^4  \alpha^2 c  
\approx 3.25 \times 10^{-18} s ; \ \
t_{_C} \propto \hbar^3
\tag{22}
\end{equation}
where $\alpha = e^2 / \hbar c \approx 1/137$
is the fine structure constant.
(It is worth noting that a ``$U$-scale''
$\theta_{_U} = $ $( t_{_C} / t_{_{U}} )^{1/4}
\approx 1.65 \times 10^{-9}$,
where $t_{_{U}} \approx 4.4 \times 10^{17} s$
is the age of the universe,
comes close to the current CMB temperature,
$\theta \approx$ $ 0.534 \times 10^{-9}$.)
In dimensionless terms, Eq. (19)
for Thompson limit can now be rewritten as:
\begin{equation}
( d \mu / d \tau )_{_{Th}} = f_{_{Th}} = - \theta^4 \mu \gamma
\tag{23}
\end{equation}
This approximation is valid for $K_C \ll 1$
(hence $\theta \ll 1$),
and it is still good for relativistic
case, $\mu \sim \gamma \gg 1$,
as long as $\mu \ll \theta^{-1}$.
(Note that for electrons, 
$ \theta = 1$ corresponds to $T \approx 0.6 \times 10^{10} K$).
Eq (23) is readily solved for
$\mu (\tau )$; with an initial condition
$\mu = \mu_{_0}$ at $\tau = 0$, we have
\begin{equation}
\mu ( \tau ) = 1 / {\sinh ( \tau \theta^4 + \delta_0 ) } ;
\tag{24}
\end{equation}
where $\delta_0 = ln [(1 + \gamma_0 )/ \mu_0]$.
At $\mu \ll 1$ Eq. (24)
reduces to $\mu \propto \exp ( - t / t_{_{TR}} )$,
while in relativistic case, $\mu \gg 1$, -- to
$\mu \approx \mu_0 ( t / t_{_{TR}} + 1 )^{-1}$;
a time scale here is $t_{_{TR}}= t_{_C} / \gamma_0 \theta^4$.
This scale may vary tremendously
even for $\mu_0 \lesssim 1$ --
from $0.4 \times 10^{-10} s = 40 ps$ for 
the temperature $\sim 10^8 K$ ($\theta \sim 1.7 \times 10^{-2} $)
below lab-nuclear fusion, to $0.4 \times 10^{20} s$ --
for the current epoch CMB
($\theta \approx 0.534 \times 10^{-9}$),
which is $10^2$ times longer
than the age of the universe, $t_{_U}$.

For a frequency-dependent
cross-section $\sigma_{_{MT}} ( \epsilon )$,
where $\epsilon = \hbar \omega / m_0 c^2$,
Eq. (17) for a blackbody radiation (21),
can be readily reduced to a single integral by
using the {\emph {dilogarithm}} function,
$Li_2 ( z ) = - \int_0^z ln ( 1 - t ) dt / t$ [45]:
\begin{equation}
f = - \frac {45} {8 \pi^4} 
\frac {\theta^4} { ( \mu \gamma ) ^2 }
\int_0^{\infty} \frac {\sigma_{_{MT}} ( x \theta / \gamma )} {\sigma_0}
( S^- + S^+ ) dx ; \ \ \ \
\tag{25}
\end{equation}
where
\vspace{-0.1in}
\begin{equation}
S^{\pm} = \pm x Li_2 \left[e^{-x ( 1 \pm \beta )}\right] -
x^2 \beta \ln \left[ 1 - e^{-x ( 1 \pm \beta )}\right]
\notag
\end{equation}
and $\beta = \mu / \gamma$.
In particular, for a low-relativistic
motion, $\beta \sim \mu \ll 1$,
but arbitrary high temperature $\theta$,
Eq. (25) is further reduced to
$f = - \mu \theta^4 \Theta ( \theta )$, where
\begin{equation}
\Theta ( \theta ) = \frac {15} {16 \pi^4} 
\int_0^{\infty} 
\frac {\sigma_{_{MT}} ( x \theta ) x^4} 
{\sigma_{_0} \ sinh^2 ( x / 2 ) } dx ; \ \ \
\Theta ( 0 ) = 1
\tag{26}
\end{equation}
For arbitrary $\mu$, a specific case
of electron is considered in Sect. VIII below,
but we need first to determine a QED-related 
energy dependence of scattering \& momentum-transfer
cross-sections for electron
in the next Section.
\vspace{-0.2in}
\section{QED scattering \& momentum-transfer 
cross-sections for electron}
In the limit of a low-energy photons,  $ \epsilon \ll 1$,
their scattering by a charged particle
is described by an energy-independent 
Thompson cross-section 
$\sigma_0$, Eq. (20).
Yet a cross-section $\sigma$ becomes energy-dependent even 
at sub-relativistic energies,
which in the case of electrons/leptons is
due to quantum Compton scattering
$via$ Klein-Nishina theory providing
an exact solution for $\sigma = \sigma_{_{KN}}$ for any $ \epsilon$,
good to the first degree in $\alpha$.
The differential cross-section 
$d \sigma_{_{KN}} / dO$ in that case is [39,40]:
\begin{equation}
d \sigma_{_{KN}} / {dO} =
( 3 \sigma_{_0} / 16 \pi ) R^2 
\left( R + R^{-1} - \sin^2 \psi \right)
\tag{27}
\end{equation}
with $R = R ( \epsilon , \psi )$ as in Eq. (8).
Using Eq. (27) in Eq. (13) we get a 
full Klein-Nishina cross-section [39,40]:
\begin{equation}
\frac{\sigma_{_{KN}} ( \epsilon )}{\sigma_{_0}} =
\frac{3}{8 \epsilon} \left[ \left( 1 - \frac{2}{\epsilon} - \frac{2}{\epsilon^2}
\right) \ln ( 1 + 2 \epsilon ) + \right.
\notag
\end{equation}
\vspace{-.2in}
\begin{equation}
\left.
\frac{1}{2} + \frac{4}{\epsilon} - \frac{1}{2 ( 1 + 2 \epsilon )^2 }
\right]~~~~with~~~~
\epsilon \equiv \frac{\hbar \omega}{m_{_0} c^2}
\tag{28}
\end{equation}
In the ``cold'' and ``hot'' limits we have respectively,
${\sigma_{_{KN}}}/{\sigma_{_0}} \approx 1 - 2 \epsilon$
at $\epsilon \ll 1$; 
and $ {\sigma_{_{KN}}}/{\sigma_{_0}} \approx$ $({3}/{8 \epsilon})
[ \ln ( 2 \epsilon ) + 1/2 ]$ at $\epsilon \gg 1$.
Using now Eq. (8) in Eq. (14), we have
\begin{equation}
\frac{\sigma_{_R}}{\sigma_{_0}} = 
\frac{3}{8} \int_{-1}^1 \tilde{R^2}
[ \tilde{R^2} - ( 1 - \zeta^2 ) \tilde{R} + 1 ]
\zeta d \zeta
\tag{29}
\end{equation}
where $\tilde{R} = [ 1 + \epsilon ( 1 - \zeta ) ]^{-1}$.
Its integration yields:
\begin{equation}
\frac{\sigma_{_R} ( \epsilon )}{\sigma_{_0}} =
\frac{3}{8} \left\{
\frac{ 2 \epsilon ( 1 - 2 \epsilon ) }{3 ( 1 + 2 \epsilon )^3 } +
\left[ \frac{2 ( 1 + \epsilon )}{ \epsilon ( 1 + 2 \epsilon ) }
- \frac{\ln ( 1 + 2 \epsilon )}{\epsilon^2} \right] +  
\right.
\notag
\end{equation}
\begin{equation}
\left.
\left\{
\frac{2 ( 2 + 4 \epsilon +  \epsilon^2 )}{\epsilon ( 1 + 2 \epsilon )^2} -
\frac{3 ( 1 + \epsilon )}{\epsilon^3}
\left[\frac{2 ( 1 + \epsilon )}{1 + 2 \epsilon } -
\frac{\ln ( 1 + 2 \epsilon )}{\epsilon}
\right]
\right\}
\right\}
\tag{30}
\end{equation}
As mentioned already, $\sigma_{_R} ( \epsilon = 0 ) = 0$.
The three terms within the ``outer''
brackets $\{...\}$ in Eq. (30) are grouped 
to have each one of them to also zero out at $\epsilon = 0$. 
In the ``cold'' and ``hot'' limits we have respectively
${\sigma_{_R}}/{\sigma_{_0}} \approx 6 \epsilon / 5$
at $\epsilon \ll 1$,
and $ {\sigma_{_R}}/{\sigma_{_0}} \approx$ $({1}/{2 \epsilon})
[ 1 - 3 \ln ( 1 + 2 \epsilon ) / 4 \epsilon ]$ at $\epsilon \gg 1$,
so that the term $\sigma_{_R}$ in the 
photon$\leftrightarrow$electron momentum
transfer can be neglected at $\epsilon , \epsilon^{-1} \ll 1$.
It can be shown that 
$\sigma_{_{KN}} >$ $ \sigma_{_R} > 0$ 
everywhere in $\epsilon \in ( 0 , \infty )$,
so that total cross-section 
$\sigma_{_{MT}} = \sigma_{_{KN}} - \sigma_{_R}$
in Eqs. (17) and (18) is always positively defined.
All the $\sigma$'s spectral profiles are depicted at Fig. 1 
showing that $\sigma_{_R} ( \epsilon )$
peaks as $\sigma_{_R} / \sigma_{_0} \approx 0.1442$
at $\epsilon \approx 0.543$ ($\hbar \omega \approx 277.5 \ KeV$).
At that point $\sigma_{_{KN}} \approx 0.55 \sigma_{_0}$,
and $\sigma_{_{MT}} \approx 0.4 \sigma_{_0}$,
i. e. $\sigma_{_R} / \sigma_{_{KN}} \approx 0.26 \gg \alpha$.
Thus, strictly speaking,
$\sigma_{_R}$ should not be neglected within
KN-theory using $O ( \alpha )$ terms,
at least around $\epsilon \sim 1 $.
Yet in reality, it makes little difference
when calculating $f ( \mu, \theta )$
in the entire momentum span,
$\mu \in ( 0 , \mu_{_{Pl}} )$,
where $\mu_{_{Pl}} =$ $k_B T_{Pl} / 
m_0 c^2 \approx 2.4 \times 10^{22}$
is the highest momentum in the universe
related to the Planck temperature
$T_{Pl} \approx 1.417 \times 10^{32} K$, 
see below.
\begin{figure}
\centerline{\includegraphics[angle=270,width=3in]{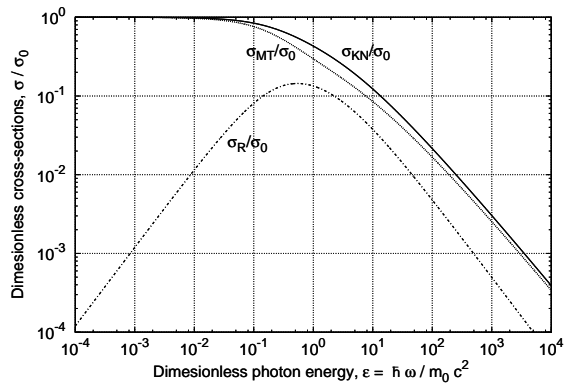}}
\caption{Dimensionless integrated cross-sections
of photons scattering by an electron:
Klein-Nishina $\sigma_{_{KN}} / \sigma_{_0}$,
``projection'' term $\sigma_{_R} / \sigma_{_0}$,
and momentum-transfer $\sigma_{_{MT}} / \sigma_{_0} = 
( \sigma_{_{KN}} -$ $ \sigma_{_R} ) / \sigma_{_0}$,
$vs$ dimensionless energy of incident photons.  }
\label{fig1}
\end{figure}
\begin{figure}
\centerline{\includegraphics[angle=270,width=3in]{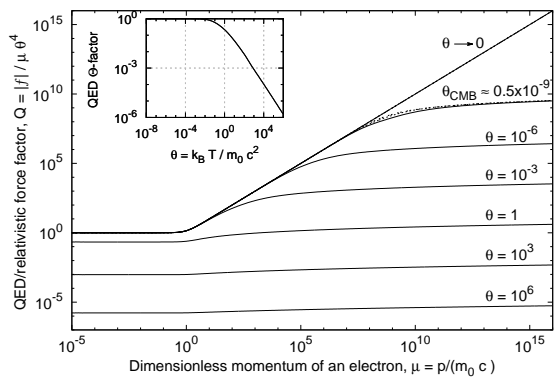}}
\caption
{QED/relativistic factor, $Q = |f| / \mu \theta^4$
$vs$ momentum, $\mu$,
for various temperatures, $\theta$.
Three curves slightly divergent around
$\mu \sim \mu_{_C}$ in the case
$\theta = \theta_{_{CMB}}$ correspond
to numeric integration of Eq. (17)
with $\sigma_{_{MT}} = \sigma_{_{KN}} - \sigma_{_R}$
(upper), $\sigma_{_{MT}} \approx \sigma_{_{KN}}$
(middle), and analytic Eq. (32) (lower).
Similar curves for $\theta = 10^6$ coincide
to the line width.
Inset: a QED factor $\Theta$, Eq. (27), 
for non-relativistic motion $vs$ $\theta$.
}
\label{fig2}
\end{figure}
\vspace{-0.1in}
\section{QED blackbody light pressure}
Eqs. (28),(30) together with Eq. (24)-(26)  allow for specific
investigation of the light pressure 
by Planck radiation on an electron.
The force $f$ $vs$ $\mu$ in the entire momentum span,
$\mu \in ( 0 , \mu_{_{Pl}} )$, and for various $\theta$,
from $\theta \sim 10^{-9}$ to $10^{6}$,
was numerically evaluated using Eq. (25)
and depicted in Fig. 2 for the relativistic/QED 
factor $Q ( \mu , \theta ) = - f / \mu \theta^4 $,
where $- \mu \theta^4 $ is the Thompson non-relativistic
light pressure, Eq. (23), for $\gamma = 1$;
for $\mu , \theta \ll 1$, $Q = 1$.
In ultra-relativistic QED, or Compton
domain, based on the 
behavior of $\sigma_{_{KN}} ( \omega )$
at $\omega \rightarrow \infty$, Eq. (28),
the asymptotics of $Q$ at
$\mu \gg max ( 1, \theta^{-1} )$ can be shown to be
$Q \propto [ \ln ( \theta \mu ) + O (1) ] /  \theta$.
Using these numerical and asymptotic results, 
it would be greatly
beneficial for further analysis to have their
good analytic interpolation/approximation.
Amazingly, this task is perfectly served by a
remarkably simple formula good for the entire
span $\mu \in ( 0 , \mu_{Pl} )$ and
$\theta \in ( 0 , \theta_{Pl} )$:
\begin{equation}
f_{M} ( \mu , \theta ) = - \mu \theta^3 
ln \left( 1 + K_C \right) / q ; \ \ \ \
K_C = {\gamma} \theta q
\tag{31}
\end{equation}
where $q = 10$ is a fitting parameter.
Eq. (31) makes better than a few percents fit
to the numerics over the entire span of momentum
but a small area near $K_C \sim 1$, Fig. 2,
where they are still very close;
it can be viewed as a benchmark for 
any other possible approximations.
In the limit $K_C \ll 1$, 
Eq. (31) is reduced to Eq. (23).
For low-relativistic motion, $\mu \ll 1$, $\gamma \approx 1$,
yet arbitrary temperature,
the factor $\Theta (\theta )$ in Eq. (26),
shown at the inset in Fig. 2,
is now approximated using Eq. (31) as 
\begin{equation}
\Theta = ln ( 1 + \theta q ) / \theta q ; \ \ \ \
( \Theta \approx 1 \ \ \ at \ \ \ \ \theta q \ll 1 )
\tag{32}
\end{equation}
\vspace{-0.2in}
\section{Kinetics of density distribution 
near to and far from equilibrium}
Since $F \equiv dp / dt$, Eq. (15),
or $f = d \mu / d \tau$ [see e. g. Eq. (23)],
the averaged dynamics of electron motion, $\mu ( \tau )$,
for $\theta = const$ is implicitly 
described as
$\tau = \int d \mu / f ( \mu , \theta )$.
It is readily integrated 
in the case of Thompson scattering, $K_C \ll 1$ 
yielding an explicit function $\mu ( \tau )$, Eq. (24),
whereas in general case,
especially for the transition
from Compton domain, $K_C \gg 1$
to the Thompson domain,
the timeline becomes more complicated,
yet still analytically solvable
using another approximate model,
which is very close to the Eq. (31)
in the energy span covering the entire
Compton and most of the Thompson domains [41].
In the context of this paper,
an important issue is the relaxation time $\tau_{rlx}$
(or rate $\tau_{rlx}^{-1}$)
$vs$ temperature $T$ of the electron 
distribution to its equilibrium state at a given $T$,
and the momentum $\mu$ of a non-equilibrium electron.
This problem is best handled by using a Fokker-Planck equation
for the diffusion in the momentum space [46].
To that end, we consider
a distribution function, $ g^{(e)} ( \mu , \tau )$
of electrons defined here as the number of electrons per 
element of solid angle $d O$, element of momentum, $d \mu$,
within a unity of coordinate space,
and a density number,
$\rho^{(e)} ( \mu , \tau ) = $ $4 \pi \mu^2$ $ g^{(e)}$.
Note that in application to cosmology,
whereby one needs to consider the expanding space/universe,
these functions reflect the distribution within
the $expanding$ unity of coordinate space,
i. e. in the spatial ``unity box''
expanding at the same rate as the Universe.
Assuming then that
(a) the electron distribution is isotropic, same as CMB,
(b) the total number of electrons, 
in the unity of momentum space
and the expanding unity of coordinate space
is approximately invariant,
$\int_0^{\infty} \rho^{(e)} d \mu = N_{\Sigma} = inv$,
and (c) the thermal equilibrium of 
a relativistic gas at any $\theta$
is  due to the Maxwell-J\"{u}ttner (MJ) distribution [47,48],
\begin{equation}
g_{_{MJ}}^{(e)} \propto  e^{- \gamma /\theta}
[\theta K_2 ( 1 / \theta ) ]^{-1}
\tag{33}
\end{equation}
where $K_2$ is the modified Bessel function of the second order,
with MJ being a relativistic generalization 
of the Maxwell-Boltzmann distribution,
\begin{equation}
g_{_{MB}}^{(e)} \propto  e^{- \mu^2 / 2 \theta} \theta^{- 3/2}
\tag{34}
\end{equation}
we found a Fokker-Planck equation 
for the distribution function $g^{(e)} ( \mu , \tau )$,
in terms of dimensionless momentum $\mu$, factor $\gamma$,
time $\tau$, temperature $\theta$, and force $f$, as
\vspace{-.05in}
\begin{equation}
\frac {\partial [ \mu^2 g^{(e)}] } {\partial \tau} +
\frac {\partial } {\partial \mu}
\left\{ \mu^2 f ( \mu , \tau ) \left[ g^{(e)} +
\theta ( \tau ) \frac {\gamma } {\mu}
\frac {\partial g^{(e)}} {\partial \mu} \right] \right\} = 0
\tag{35}
\end{equation}
This equation is valid for time-dependant $\theta$
and for the expanding universe 
For our purposes here we will consider 
only the case $\theta = const = \theta_{eq}$
(and neglect the universe expansion),
since the time involved is mush shorter
than the age of universe for the effects
of interest here.
Eq. (35) is solved fully analytically for the
non-relativistic case, $\mu \ll 1$,
whereby any initial distribution function $ g^{(e)} ( \tau = 0 )$
is decomposed into Gaussian components, Eq. (34),
each of which has its specific initial
temperature, $\theta_{in}^{(e)}$ at $\tau = 0$;
in time, their distribution functions remain Gaussian, 
with their time-dependant effective temperature being as:
\begin{equation}
\theta^{(e)} ( \tau ) = \theta_{eq} + 
( \theta_{in}^{(e)} - \theta_{eq} ) e^{ - \tau / \tau_{rlx} } ;
\ \ \ 
\tau_{rlx} = 1 / 2 \theta_{eq}^4
\tag{36}
\end{equation}
i. e. $t_{rlx} = t_{TR}/2 = t_C / 2 \theta_{eq}^4$
(see eq. (24) and explanations therein).
Eq. (36) still holds approximately
for the entire Thompson domain
for initial conditions close to the equilibrium
(be reminded that the condition
$K_C \ll 1$ automatically means
$\theta_{eq} \ll 1$,
while proximity to the equilibrium
-- that $\gamma_{in} = O ( 1 )$ ).
In general case of arbitrary initial conditions
the relaxation rate becomes
\begin{equation}
\tau_{rlx}^{-1} \approx q^{-1} [ 2 \theta_{eq}^3 ln ( 1 + 
q \theta_{eq} \gamma_{in} ) ] ; \ \ \
\tag{37}
\end{equation}
where $ \gamma_{in}$ is the energy
at the peak of initial density distribution.
If it was $MJ$ density distribution
$\rho_{MJ}^{(e)} \propto \mu^2 g_{MJ}^{(e)}$,
of the initial temperature $\theta_{in}$,
then  $ \gamma_{in} = [ 1+ 
2 \theta_{in} ( \theta_{in} + \sqrt{1 + \theta_{in}^2 } \ ) ]^{1/2}$,
and $t_{rlx} = t_C \tau_{rlx}$.
For $\theta_{in} \ll 1$, it coincides with Eq. (36),
and for $\theta_{in}^2 \gg 1$ we have
\begin{equation}
\tau_{rlx} \approx q [ 2 \theta_{eq}^3 
ln ( 2 q \theta_{eq} \theta_{in}  ) ]^{-1} .
\tag{38}
\end{equation}
It is instructive to look at a few examples of interest.
For the current CMB,
the relaxation time to the equilibrium
will exceed the age of universe by $\sim 10^2$,
see. Eq. (23) and related discussion,
which makes the thermalization here almost irrelevant.
With an easily lab-accessible $T = 1300 K$ 
($\theta = $ $2.2  \times 10^{-7}$),
we have $t_{rlx} \approx 0.7 \times 10^9 s$,
which is still unrealistic in practical terms.
At the "Compton threshold", $K_C = 1$, $\theta = 0.1$,
we have $t_{rlx} \approx 2.3 \times 10^{-14} s = 23 fs$.
With $T \sim $ $ 10^{10} K $ 
($\theta \approx 1.7$) required for 
the controlled nuclear fusion [49,50],
we have $t_{rlx} \sim
10^{-18} s = 1 as$ for $\gamma_{in} \sim 1$.
If $\gamma_{in} \gg 1$, 
the process is getting even faster,
and the radiation might act to almost
instantly deplete coherency
of e. g. an electron beam used as a plasma diagnostic tool.
For example, if its energy is $50 MeV$, a Compton
factor in Eq. (37) is 
$K_C = q \theta_{eq} \gamma_{in} \approx 1,700$,
with $t_{rlx} \approx 0.4 as$,
so that the starting rate
of e-beam thermalization is much faster.

Having in mind plasma nuclear fusion,
it would be of substantial interest
to see how the light pressure-induced 
damping may affect plasma oscillations
in high-density, high-temperature plasmas.
Those many-body excitations 
seem to preclude a single-electron
light pressure, Eq. (31), 
from playing a significant role in plasma relaxation.
However, the relaxation time 
$t_{rlx}$ due to light pressure
does become a major player as long as
it gets faster than relaxation in
other channels of thermalization 
such as collision with similar or other
species (e. g. electrons and protons), $t_{cls}$, 
to dominate in a total relaxation rate 
$t_{\Sigma}^{-1} = t_{rlx}^{-1} + t_{cls}^{-1}$
if $t_{rlx} \ll t_{cls}$, hence
$t_{\Sigma} \approx t_{rlx}$.
Yet the most important effect here
is that the time $t_{rlx}$ may get even
much shorter than a plasma oscillation period,
$t_{rlx} \omega_{pl} \ll 1$,
in which case those oscillations 
would be damped or even extinguished.
Using in rough approximation
a standard equation for plasma
frequency $\omega_{pl}$ $vs$ the number density of 
electrons, $N_e$, we get the condition
on $N_e$ to have plasma oscillations
strongly damped 
\begin{equation}
N_e \ll N_{cr} = \frac {m_0 / \pi e^2} 
{( t_{_C} \tau_{rlx} )^2} \approx 
\frac {1.2 \times 10^{26}}
{\tau_{rlx}^2} cm^{-3}
\tag{39}
\end{equation}
up to $\omega \sim \omega_{dmp} = t_{rlx}^{-1}$,
where $N_{cr}$ is a critical number density of electrons
below which the light pressure suppresses plasma oscillations.
For $T \sim 10^{10} K$
($\theta = 1.7$) required for the nuclear fusion [49,50],
presuming near-equilibrium, $ \theta_{eq} \approx \theta_{in}$,
and using Compton domain Eq. (38),
we have $N_{cr} \approx 2 \times 10^{27} cm^{-3}$,
which exceeds the number density
of the interior of most of the stars [29-31], 
the Sun incuding, thus making plasma 
oscillations of electrons completely extinguished.
Even with an order of magnitude lower temperature 
$T \sim 10^{9} K$,
using Thompson formula (36) for $\tau_{rlx}$
in Eq. (39), we get $N_{cr} \approx 3.4 \times 10^{20} cm^{-3}$,
which is still much higher than any conceivable
lab plasma density.

Notice that far from equilibrium,
with $\theta_{in} \gg \theta_{eq}$ in Eqs. (37), (38),
the thermalization rate, $\tau_{rlx}^{-1}$,
as well as the "friction coefficient",
$f / \mu$ in Eq. (31),
still increase with energy $\gamma_{in}$ 
(which translates into photon energy $\epsilon$
increase in $P$-frame)
in the Compton domain, $K_C \gg 1$,
for a fixed temperature,
while cross-section $\sigma ( \epsilon )$ decreases.
The explanation of this is that while 
the $\sigma ( \epsilon )$ is indeed slowly
receding with photon energy
(as $\sim 1 / ln (\epsilon )$, Eqs. (28) and (30)), 
each act of Inverse Compton Scattering
(ICS) gets much more ``quantum efficient''
since then a low-energy photon scattered from a
high-energy electron gets a huge boost
by accruing up to almost full energy of the electron.
The peak gain is reached in a ``head-on'' collision, 
when a photon is exactly back-scattered. 
Based on the Compton scattering formula, Eq. (8),
and Doppler effect, Eq. (3), that enter the final
formula, Eq. (17),
the maximum scattered photon energy in the $L$-frame is 
\begin{equation}
\epsilon_{sc} \equiv \hbar \omega_{sc} / m_0 c^2 =
{\epsilon_{in} ( \gamma +  \mu )} / 
( \gamma -  \mu + 2 \epsilon_{in} )
\tag{40}
\end{equation}
where $\epsilon_{in}$ is the incident photon energy.
For high-energy electrons,
$\gamma \approx \mu \gg 1$, and low-energy
incident photons, $\epsilon_{in} \ll 1$,
the maximum quantum efficiency of ICS
defined as the ratio of scattered photon
energy to that of an incident electron,
$\eta \equiv {\epsilon_{sc}} / {\gamma}$ ,
is then $\eta \approx$ 
$[ 1 + $ $( 4 \gamma  \epsilon_{in} )^{-1} ]^{-1}$,
and in the sub-QED domain, 
$\gamma \ll 1 / \epsilon_{in}$, we have
$\eta \approx  4 \gamma  \epsilon_{in} \ll 1$,
hence small loss of electron energy per collision.
However, in QED Compton domain, 
$4 \gamma  \epsilon_{in} \gg 1$,
we have $\eta \approx 1 - ( 4 \gamma  \epsilon_{in} )^{-1} \sim 1$,
i. e. an electron passes great part of its energy to a scattered photon.
With considerable probability an electron
jumps then in one collision to the Compton threshold,
$\mu \rightarrow \mu_{sc} \sim 1/\theta q$.

These effects make it of special interest
to look into the kinetics of highly 
relativistic electrons with their energy far exceeding
that of equilibrium.
Since in such a case the system
remains far from the equilibrium
during the evolution, the last term
in Eq. (35) can be omitted,
and in terms of density
$\rho^{(e)} \propto \mu^2 g^{(e)} $ it can be reduced to
\vspace{-.05in}
\begin{equation}
{\partial \rho^{(e)}} / {\partial \tau} +
{\partial [ f \rho^{(e)} ] } / {\partial \mu} = 0
\tag{41}
\end{equation}
which is essentially a continuity-like equation and
is fully integrable; its general solution can be shown to be
\vspace{-.05in}
\begin{equation}
\rho^{(e)} = {\Phi ( \xi - \tau )} / {f ( \mu )} ,~~~with~~~
\xi = \textstyle \int  d \mu / f
\tag{42}
\end{equation}
where $\Phi (s)$ is an arbitrary function of $s$;
here it is defined by initial conditions,
e. g. the MJ-distribution with $\theta_{in} \gg 1$.
Resulting evolution of the
spectra of high-$T$ sources in the relict radiation
environment reveals their drastic
transformation, e. g.  formation of narrow spectral
lines in the cosmic electron spectrum,   
and development of a "frozen non-equilibrium" state due
to very low rate of electron momentum decay
in the Thompson domain.
The ramifications of these effects 
in astrophysics and cosmology
will be discussed by us elsewhere [41].
\vspace {-.2in}
\section{Conclusion}
\vspace {-.1in}
In conclusion, we developed a theory for
the light pressure on particles,
in particular electrons, by isotropic radiation,
in particular blackbody/Planck radiation,
that covers the entire span
of energies/momenta up to the Planck energy
by using, in the case of electron,
the QED Klein-Nishina theory for electron-photon
cross-section scattering.
We also analyzed the kinetics 
of electron relaxation into equilibrium 
by using relativistic Fokker-Planck equation
for temporal evolution of 
electron spectra of high-$T$ sources,
revealing dramatic difference between
classical and QED electron relaxation rates
that may result in a host of new effects. 
We showed as well that a light pressure-induced
damping may completely extinguish
plasma oscillations of electrons at the
temperatures approaching nuclear fusion.

The author is grateful to B. Y. Zeldovich
for insightful discussions and to anonymous
referees for very thoughtful and helpful comments.
\vspace {-.01in}

\end{document}